# Detailed Review of Cloud based Mobile application for the stroke patient

Balagopal Ramdurai

*Masters of Business Administration- ITM*
*Dubai- United Arab Emirates*

**Abstract:** In the current years, due to the significant developments in technologies in almost every domain, the standard of living has been improved. Emergence of latest innovations, advanced machinery and equipment especially in the healthcare domain, have simplified the diagonalizing process to a wide extent. Smart techniques employed in medical applications resolved the detection and rectification of various diseases. This work reviews the cloud based mobile application for stroke patients. The prime goal of this research is to study the challenges and necessary measures to be implemented for the rehabilitation of patients on post-stroke. Furthermore, the existing cloud-based services and the techniques to be modified for the improvement in the patients' health status need to be explored.

**Keywords:** *stroke, cloud computing, healthcare, mobile application*.
.

## I. INTRODUCTION

In recent years, due to the rapid advancements in technologies, there has been a lot of improvement in the medical sector. Lots of innovations in the medical sciences have led to the emergence of advanced machinery and equipment simplifying the process to a huge extent. In spite of the advances in the healthcare sector, the problem of figuring out the root cause of newly emerging diseases is still persisting (Lou et. al., 2017). However, the trending techniques like cloud computing, block chain. artificial intelligence etc. employed in the healthcare sector, have resolved this issue to a certain extent. Research for the detection and diagnosis of several diseases is progressing in the medical sector.

Stroke is a chronic disease and is considered to be the major cause of the public health crisis. The population of stroke patients has been rising every year and the death rate is also growing simultaneously. It has been estimated by researchers that someone dies every 4 mins, due to stroke (Zhou et. al., 2018). In the current scenario, stroke is considered to be the eminent source of adult disability. People diagnosed with stroke either suffer with prolonged disability or get affected by cognitive, physical, social functioning. Hence optimal rehabilitation is important in this regard (Munich et. al., 2017). The stroke patients need to be practically and emotionally strengthened for health improvement. Additionally, awareness of social isolation risk is very essential among rehabilitation experts (Karaca et. al., 2019). About 11.13% of deaths are said to have occurred due to stroke and is considered to be the second largest source of death. Due to the advancements in mobile healthcare, there has been huge improvement in patient care leading to patient's health status.

### A. Cloud based services:

With rapid technology advancement cloud becomes an important pillar for any solution. The application for improvising patient care needs to be stored in the cloud for scalability and complete availability with security. The data is used for reading and analysing the pattern in stroke patients for improving recovery rates.

#### a) Medical Distributed Utilization of Services & Applications (MEDUSA)

This technique is based on a cloud platform. It is capable of advanced medical data processing, quick information exchange, smart decision making, remote collaboration etc. As a result, it simplifies the process of acquiring and retrieval of medical data through its advanced features. This platform is highly beneficial in the healthcare sector. Data monitoring of stroke patients and the necessary measures to be implemented for rehabilitation can be easily achieved through this platform (Barros et al., 2015).

#### b) Clinical Performance Monitoring Application (CPMA)

The prime motive of this application was to monitor the clinical data of stroke patients for rehabilitation. The application was designed using three phases. The phases comprised modelling, implementation and evaluation. Modelling phase involved identifying the goals, defining the clinical metrics for the analysis, determining the data resources. Implementation phase involved configuring applications, application reports and mapping the clinical data to the reporting database. In the evaluation phase, feedback obtained from user experiences were analysed by the technical and medical experts for implementing suitable measures for further improvement in healthcare (Mata et. al., 2016).

## II. REVIEW OF EXISTING STUDIES

Cao et al., (2015) proposed a technique for reducing the stroke through fog computing. The proposed U-Fall technique performed better in terms of sensitivity and specificity. About 75% of specificity and 88% of sensitivity were achieved. The proposed algorithm was implemented using cost-effective devices. However, the need for further improvement in specificity was illustrated. The requirement of huge datasets pertaining to biomedical applications and their experimentation for enhancing the systems' efficacy were highlighted. Further investigation regarding the implementation of the proposed algorithm in the real-time applications is required.

Seoet. al., (2015) proposed a technique for validating the effectiveness of mobile application in determining the vascular threats in stroke patients. The analysis was conducted using clinical data of 6 months. The study revealed that dyslipidaemias, hypertension and diabetes were the basis of anticipation of stroke in patients. The approach was successful in testing the potentiality of the mobile application. However, the proposed technique was unable to determine the outcomes of drug adjustment. For the analysis, data of people with only 20-80 years of age were collected and neurologically deficient patients were excluded in the study. Complications in the data registration process of users was the main drawback of this approach.

Andrew et. al., (2017) proposed a mobile application for analysing the clinical metrics and its effects in stroke patients. About 4953 cases were used for the study. From the results, it was inferred that out of 4953 cases, 1173 cases were considered as erroneous entries and 1190 cases for inclusion criteria. Real-time data collected through smartphones were highly beneficial in analysing acute strokes. It was estimated that about 9.7% of age-related cases and 7.4% of sex related cases were missing during the data analysis. Usage of huge sample size enhanced the effectiveness of the study. However, the inability of the technique to address the endovascular intervention issues was considered to be the drawback of this study.

Richardson et al., (2019) conducted a survey of wearables for stroke by acquiring data without the intervention of the user. Feedbacks were produced by the application of data mining techniques. For the analysis, data was collected from the public sources. Videos of patients diagnosed with Aphasia/Dysarthria were used as datasets. The dataset included 16 patients diagnosed with stroke of which 8 were categorised as Aphasia and remaining were categorized as Dysarthria. The desired speech samples were extracted from the videos and the irrelevant speech samples were eliminated. For the experimental purpose, data was split into 4 test sets and 4 even sets. The proposed technique was successful in classifying the speech samples for stroke. This study indicated the feasibility of using speech samples in the unconstrained area. However, further research in analysing the walking and typing data acquired from sensors needs to be addressed.

Piranet. al., (2019) designed mobile applications for analysing the performance of stroke patients. The study used 18 exclusion criteria for the research. Results estimated that among 30,132 applications, 2.7% were considered to be highly beneficial and 8.7% were modelled for caregivers and survivors of stroke. The applications designed were successful in providing the visual attention and language therapy, estimation of stroke risk, patients communication, rehabilitation, identification of acute stroke and atrial fibrillation services. About 769 applications were designed for detecting other diseases. The proposed scheme effectively solved the problems pertaining to communication and language. However, the study reviewed only Apple iTunes applications excluding the reminder application pertaining to medication. Sleep apnoea risk factors were not considered included for the analysis.

Hodaet. al., (2015) proposed a technique for rehabilitation of patients diagnosed with stroke and their recovery on post-stroke. To improve the motor functions of patients on post-stroke was the prime aim of this work. Analysis was conducted using the data of 45 healthy people and 3 people diagnosed with stroke. Kinematic patterns of patients diagnosed with stroke and healthy people were compared using Dynamic time warping and optical alignment methods. Recovery performance of patients on post-stroke were analysed using Auto-Regressive Integrated Moving Average model (ARIMA). Performance of ten weeks was estimated by computing the four sessions average. But analysis used only the first eight weeks of data and the last 2 weeks data for forecasting. Testing the Jerkiness in the curves demonstrated the patients' upper limb instability. Post rehabilitation, feedback regarding the patients' experience were collected for further improvement. On evaluating the feedback, it was noted that the proposed technique satisfied the patients to a huge extent.

Further, assessment of patients' health status was simplified due to the updated information provided by the system based on feedback. Improvement in the upper limbs movement and control during the training period were demonstrated in the results. The approach was successful in testing the patients. For patient numbered 1 and 3 an error lesser than 2.0% was achieved and 10.35% was achieved for patient number 3. Rehabilitation performance was effectively studied using DTW algorithm. High precision rate was achieved in the recovery process of patients using ARIMA. However, the need for analysis with a large number of patients should be addressed. Furthermore, analysis needs to be conducted using different kinematic variables for accuracy enhancement.

## III. COMPARATIVE ANALYSIS

| Refs | Objectives | Applications | Techniques | Obtained results / Merits | Demerits |
|---|---|---|---|---|---|
| Ali et al., (2018) | To describe thescope,effects, challenges and applications of cloud computing in the medical sector. | Healthcare applications | Cloud computing | The various challenges, advantages and applications in the healthcare sector were illustrated. The research can be highly beneficial for developing new technologies in future. | The need for further research pertaining to the privacy of patient data was illustrated. The techniques for proper decision making based on the updated data needs to be proposed. |
| García et al., (2019) | To monitor and determine the cerebral stroke through cloud technique. | Healthcare sector | Android studio developmen-t, Android API 26 SDK | Results concluded that among 90 tests, accurate results were obtained from 16 tests and 74 tests performed inaccurately. About 239,496 ms of time was required for conducting the complete test. The proposed technique was successful in detecting the symptoms of stroke. | Further research in detecting and analyzing the symptoms of distinct diseases were illustrated. |
| Chang et al., (2018) | To develop an application based on mobile phones for determining the efficiency of self-assessmen-t for patients diagnosed with stroke. | Healthcare sector | modified Rankin scale (mRS), activities of daily living (ADL) scale, Epidata 3.1, Chi-square | Results estimated that out of 50 patients, 49 were diagnosed with ischemic stroke and one with hemorrhagic stroke 96.8% of sensitivity and 82% of specificity were achieved. | Further analysis using large sample size needs to be undertaken. |

| | | | | mRS achieved a 0.718 kappa value. | |
|---|---|---|---|---|---|
| Requenaet al., (2019) | To develop a mobile application for creating an awareness for adopting hygienic lifestyle, simplifying the communicatio-n process with healthcare members. | Healthcare applications | Android, version 17.0 software, Fisher exact test, Mann-Whitney U test | Results concluded that 67.3% of patients were categorized under FARMALARM group and 32.7% patients under control group. It was estimated that the control group had higher levels of hemoglobin A1c (HbA1c) plasma. FARMALARM group contained a lower level of total cholesterol. Higher vascular threats in the FARMALARM group. The FARMALARM was feasible in creating stroke awareness. | Inability in randomly organizing the subjects. Lack of measures taken for prevention of stroke. Limited knowledge on smart phones usage in elderly people. |
| Fell et al., (2019) | To develop low-cost equipment for monitoring the data of caregivers, healthcare staff and patients using mobile technology. | Telemedicine | Functional Reach Test (FRT),10 Meter Walk Test (10MWT) | Proposed technique was successful in analyzing the data of 33 people. The 10MWT technique performed better in terms of clinician scores. The scores were estimated based on Spearman Rho coefficient. | Inefficiency in acquiring the data of two people out of 35. |
| Grau-Pelliceret al., (2019) | To determine the gait speed in patients diagnosed with stroke. | Medical domain | 10MWT | Walking speed of patients diagnosed with stroke was effectively estimated. Effects of Multimodal Rehabilitation Programs were studied. | Further analysis using large data size needs to be performed. Limited data was used for analysis. Assessment of data and follow-up needs to be done on a quarterly basis. |
| Gross et al., (2017) | To simplify the walking of stroke patients using a mobile robot. | Healthcare domain | Robotics | During rehabilitation, cognitive,menta-l and walking impairments were resolved. Robotic training was highly beneficial in training the patients. The effectiveness of walking in stroke patients was | Inability of proposed work in providing the training due issues relating to liability and elevator regulations. Techniques for providing the training on different floors in |

| | | | | demonstrated. | a building need to be proposed. Implementation of a robotic trainer in medical practice needs to be investigated. |
|---|---|---|---|---|---|
| Martins et al., (2019) | To evaluate the mobile application in the stroke treatment. | Medical domain | Android technique | Cost-effective and simple implementation. The application developed was successful in analyzing the performance of patients diagnosed with acute stroke. The technique simplified the physicians decision making process. | Improvement in telestroke applications. Analyzing the performance using imaging modalities. |
| Wantakaet al., (2019) | To model a technique for managing stroke patients. | Medical sector | FAST track method | FAST track technique successfully developed and managed the performance of patients diagnosed with stroke. | Effectiveness of the system needs to be improved further. |
| Sureshkum-aret al., (2015) | To design a web application for controlling the disabilities of patients on post-stroke. | Healthcare domain | Web application | Cost-effective technique. Highly benefited the stroke survivors and caregivers during their rehabilitation. Simplified the data accessing process for experts and medical staff. Feasible due to the handheld and portability features of cell phones. Content is easily accessible through laptops, tablets etc. | Further analysis using a large number of patients needs to be performed. |
| Kato et al., (2016) | To model a tele-rehabilitation system for patients on post-stroke. To improve the movement of patients during the rehabilitation process. | Healthcare sector | Kinect | Improvement in the patients movement, performance time was noted. Patient movements were tracked using 3D images. | Automatic feedback detection techniques need to be proposed by analyzing the performance. |
| Sureshkum- | To determine | Healthcare | Web application | Improvements in the | Limited data was |
| aret al., (2016) | the feasibility of the proposed web application. To figure-out the challenges encountered by the participants. | domain | | clarity of images and videos were achieved. Operational difficulties encountered during the intervention were resolved. | used for analysis. |
| Hossain et al., (2018) | To develop a technique based on cloud for rehabilitation of patients post-stroke. | Medical domain | Cloud computing, Augmented Reality | Patients gestures, rehabilitation exercises were identified effectively. Improvement in the patients recovery process was observed. | Multi-sensor-y techniques for handling huge data needs to be proposed. Analysis needs to be conducted for a large number of patients. Techniques for determining patients' stress levels needs to be designed. |

## IV. CONCLUSION

The cloud-based services employed for the healthcare domain, especially for stroke patients were reviewed. This paper discussed the existing services provided by the healthcare sector including patients who have suffered stroke. The mobile applications based on cloud for overcoming the challenges in rehabilitation period on post-stroke were explored. Further, the necessity of modification in the existing techniques for analysing the performance of stroke patients using huge data samples need to be studied. Additionally, awareness programs for educating people about the stroke effects needs to be conducted. Techniques for proper decision making after analysing the symptoms pre-stroke and protecting the privacy of data need to be developed. For better utilization of services based on smartphones by elderly people, training programs pertaining to the usage of smartphones and latest techniques embedded in the modern devices employed in the healthcare domain needs to be conducted.

## V. SUGGESTION

Telemedicine and Telehealth applications aid the doctors and patients through unified communications and hence connect people and data via the cloud. Through these applications, all stroke victims are seen by an on-call neurosurgeon through video conferencing. Mobile devices or computers are employed for accessing cloud-based technology by the doctor. 'Telestroke' an application could be used for examining the stroke patients and for conducting the entire assessment. The doctor can review radiology results and consult with other professionals as well, all via unified communications.

## VI. ACKNOWLEDGMENT

The heading of the Acknowledgment section and the References section must not be numbered.

## REFERENCES

[1] Lou, S., Carstensen, K., Jørgensen, C.R. and Nielsen, C.P., 2017. "*Stroke patients' and informal carers' experiences with life after stroke: An overview of qualitative systematic reviews*". Disability and rehabilitation, 39(3), pp.301-313.
[2] Zhou, X., Du, M. and Zhou, L., 2018. "*Use of mobile applications in post-stroke rehabilitation: a systematic review*". Topics in stroke rehabilitation, 25(7), pp.489-499.
[3] Karaca, Y., Moonis, M., Zhang, Y.D. and Gezgez, C., 2019. "*Mobile cloud computing based stroke healthcare system*". International Journal of Information Management, 45, pp.250-261.
[4] Barros, R. S., Borst, J., Kleynenberg, S., Badr, C., Ganji, R. R., de Bliek, H., ... &Olabarriaga, S. D. (2015, September). "*Remote collaboration, decision support, and on-demand medical image analysis for acute stroke care*". In European Conference on Service-Oriented and Cloud Computing (pp. 214-225). Springer, Cham.
[5] Cao, Y., Chen, S., Hou, P. and Brown, D., 2015, August. "*FAST: A fog computing assisted distributed analytics system to monitor fall for stroke mitigation*". In 2015 IEEE International Conference on Networking, Architecture and Storage (NAS) (pp. 2-11). IEEE
[6] Seo, W.K., Kang, J., Jeon, M., Lee, K., Lee, S., Kim, J.H., Oh, K. and Koh, S.B., 2015. "*Feasibility of using a mobile application for the monitoring and management of stroke-associated risk factors*". Journal of Clinical Neurology, 11(2), pp.142-148.
[7] Ali, O., Shrestha, A., Soar, J. and Wamba, S.F., 2018. "*Cloud computing-enabled healthcare opportunities, issues, and applications: A systematic review*". International Journal of Information Management, 43, pp.146-158.